\documentclass[12pt]{article}
\usepackage[dvips]{epsfig}      % lade epsfig Paket und benutze dvips Treiber
\usepackage{amsmath}            % lade amstex Paket
\usepackage{amssymb}            % lade amssymb Paket
\def\Journal#1#2#3#4{{#1} {\bf #2}, #4 (#3)}

\def\AoP{\em Ann. of Phys.}
\def\NC{\em Nuovo Cimento}

\def\NIMA{{\em Nucl. Instrum. Methods} A}
\def\NPA{{\em Nucl. Phys.} A}
\def\NPB{{\em Nucl. Phys.} B}

\def\PRL{\em Phys. Rev. Lett.}
\def\PRC{{\em Phys. Rev.} C}
\def\PRD{{\em Phys. Rev.} D}
\def\ZPA{{\em Z. Phys.} A}
\def\PR{{\em Phys. Rev.}}

\def\JPG{\em J. Phys. G: Nucl. Part. Phys.}

\def\SJNP{\em Sov. J. Nucl. Phys.}
\def\PHYSA{{\em Physica} A}
\def\bea{\begin{eqnarray}}
\def\eea{\end{eqnarray}}
\author{H. Schmieden
        \\
        {\small Institut f\"{u}r Kernphysik der 
                Johannes Gutenberg-Universit\"{a}t, 55099 Mainz, Germany}\\
       }
\title{\bf Proton Polarization in the $p(\vec e,e'\vec p)\pi^0$ \\
       Reaction and  the Measurement \\
       of Quadrupole Components \\ 
       in the N to $\Delta$ Transition}     
\begin{document}
\newcommand{\MeV}{\mbox{Me\hspace{-1pt}V}}
\newcommand{\GEn}{$G_{E,n}$}
\newcommand{\GEp}{$G_{E,p}$}
\newcommand{\GMn}{$G_{M,n}$}
\newcommand{\Den}{$D(\vec e,e'\vec n$)}
\newcommand{\Dep}{$D(\vec e,e'\vec p$)}
\newcommand{\Hep}{$H(\vec e,e'\vec p$)}
\newcommand{\Hen}{$^{3}$$\vec He(\vec e,e'n)$}
\newcommand{\snp}{S_{0+}^{*}}
\newcommand{\sem}{S_{1-}^{*}}
\newcommand{\sep}{S_{1+}^{*}}
\newcommand{\snpx}{S_{0+}}
\newcommand{\semx}{S_{1-}}
\newcommand{\sepx}{S_{1+}}
\newcommand{\enp}{E_{0+}}
\newcommand{\eem}{E_{1-}}
\newcommand{\eep}{E_{1+}}
\newcommand{\mem}{M_{1-}}
\newcommand{\mep}{M_{1+}}
\newcommand{\tcm}{\mbox{$\theta_p^{cm}$}}
\newcommand{\tlab}{\mbox{$\theta_p^{lab}$}}
\newcommand{\tw}{\mbox{$\vartheta_W$}}
\newcommand{\RL}{\mbox{$R_L$}}
\newcommand{\RLn}{\mbox{$R_L^n$}}
\newcommand{\RT}{\mbox{$R_T$}}
\newcommand{\RTn}{\mbox{$R_T^n$}}
\newcommand{\RTT}{\mbox{$R_{TT}$}}
\newcommand{\RTTn}{\mbox{$R_{TT}^n$}}
\newcommand{\RTTl}{\mbox{$R_{TT}^l$}}
\newcommand{\RTTt}{\mbox{$R_{TT}^t$}}
\newcommand{\RLT}{\mbox{$R_{LT}$}}
\newcommand{\RLTn}{\mbox{$R_{LT}^n$}}
\newcommand{\RLTl}{\mbox{$R_{LT}^l$}}
\newcommand{\RLTt}{\mbox{$R_{LT}^t$}}
\newcommand{\RLTp}{\mbox{$R_{LT'}$}}
\newcommand{\RLTpn}{\mbox{$R_{LT'}^n$}}
\newcommand{\RLTpl}{\mbox{$R_{LT'}^l$}}
\newcommand{\RLTpt}{\mbox{$R_{LT'}^t$}}
\newcommand{\RTTpl}{\mbox{$R_{TT'}^l$}}
\newcommand{\RTTpt}{\mbox{$R_{TT'}^t$}}
\newcommand{\Pn}{\mbox{$P_n$}}
\newcommand{\Pl}{\mbox{$P_l$}}
\newcommand{\Pt}{\mbox{$P_t$}}
\newcommand{\KM}{\mbox{$K_{Mott}$}}
\newcommand{\KLTp}{\mbox{$K_{LT'}$}}
\newcommand{\KLT}{\mbox{$K_{LT}$}}
\newcommand{\KTp}{\mbox{$K_{T'}$}}
\maketitle
\begin{abstract}
The recoil proton polarization in the $\pi^0$  production off the proton
with longitudinally polarized electron beam has been studied
as a means to measure quadrupole components in the N to $\Delta$
transition.
On top of the $\Delta$  resonance a  
high sensitivity to a possible Coulomb quadrupole excitation
is found in parallel kinematics.
The ratio of $S_{1+}/\mep$  multipole amplitudes 
can be determined from the ratio of the two in-scattering-plane 
recoil proton polarization components.
Avoiding the absolute measurement of the polarizations, such a ratio
allows small experimental uncertainties.
Furthermore, the electron helicity independent proton polarization component 
enables the characterization of resonant and non-resonant pieces.
% disentangle between
\end{abstract}
\section{Introduction and Motivation}

The occurrence of quadrupole components in the N to $\Delta$  transition is
within quark models related to d-state configurations in the nucleon 
and/or the $\Delta$  wavefunction \cite{Gershtein81, Isgur82}. 
They originate from details of the 
inner dynamics of the composite nucleon like a 
color hyperfine interaction in the one-gluon-exchange \cite{Glashow79}
and, therefore, are of interest for the understanding of 
the nucleon structure.
The precise measurement of the quadrupole amplitudes is a long standing 
experimental problem due to their smallness compared to the dominating 
magnetic dipole amplitude.
Only observables carrying interference terms between the large and the small 
amplitudes offer sufficient sensitivity for a reliable determination.
Appropriate interferences are accessible in $\Delta^+$-electroproduction
experiments off the proton where the resonance is tagged by its decay into
proton and $\pi^0$, and either the pion or the recoiling proton is detected 
in coincidence with the scattered electron.
Early coincidence experiments at NINA \cite{Siddle71} and 
DESY \cite{Alder72, Albrecht71, Albrecht71a}
extracted, with large experimental uncertainties, 
ratios of Coulomb quadrupole to 
magnetic dipole strength, $\sepx / \mep$, around $ - 6 \%$  over a range of 
four-momentum transfers of $0.3$  to $1.56$~(GeV/c)$^2$.
A fixed-t dispersion-relation based reanalysis \cite{Crawford71}
of older data \cite{Akerlof67, Mistretta69, Baba69}
yielded surprisingly large numbers of about $-15 \%$  
at momentum transfers down to $0.047$~(GeV/c)$^2$.
A comparatively large ratio of $(-11.0 \pm 3.7) \%$
was also obtained in a recent experiment at ELSA,
which measured the azimuthal angular distribution of the high energetic photon 
from the $\pi^0$-decay around the momentum transfer direction \cite{FK}.
All the experiments extracted the sum of resonant and 
non-resonant quadrupole components. 
A separation was achieved for the first time in 
a pion-photoproduction experiment at MAMI.
There, a linearly polarized tagged photon beam was used to determine photon 
asymmetries simultaneously for both neutral and charged 
pion production \cite{RB},
thus enabling the decomposition into isospin $1/2$  and $3/2$ channels
of the electric quadrupole amplitude, E2, at the photon point.

Further insight into the electric quadrupole admixture of the N to
$\Delta$  transition could be obtained by a precise determination of the
resonant $\sepx/\mep$  ratio as a function of four-momentum transfer.
This would constrain the spatial distribution of the electric charge
in the transition.

Polarized electron beams in
combination with polarized proton targets or recoil proton polarimetry open
possibilities for new approaches. 
The $p(\vec e,e'\vec p)\pi^0$  reaction has been examined with regard to
a measurement of the longitudinal quadrupole component
in the N to $\Delta$  transition and the separation of 
resonant and non-resonant pieces.
The next section recalls briefly the general formalism for
$\pi^0$-electroproduction and then discusses the possibilities of 
recoil polarization measurements, particularly in parallel kinematics
where the recoiling proton is detected in momentum transfer direction.
Section \ref{sec:experimentals} evaluates important experimental aspects and 
the main conclusions are summarized in section \ref{sec:conclusion}.

\section{The $p(\vec e,e'\vec p)\pi^0$  Reaction}

\begin{figure}
%% wg. hor. L\"Ucke vor ``The components ...''\vspace{1cm}
\hspace{3cm}
\epsfig{file=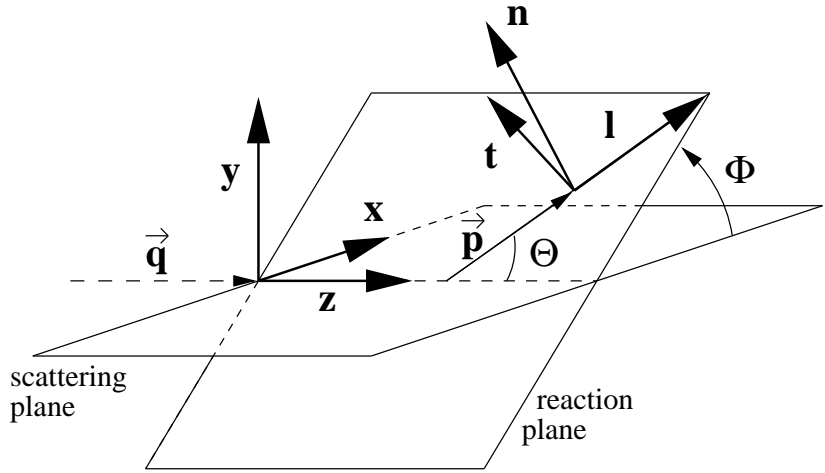}
\caption{Reference frames for the recoil proton polarization}
\label{fig:kin}
\end{figure}

Following the notation of
Raskin and Donnelly \cite{RD}, the differential cross section for 
the $p(\vec e,e'\vec p)\pi^0$  reaction can be written as
\bea \label{eq:x-sec6}
\left( \frac{d\sigma}{dE'\Omega_e\Omega_p^{cm}} \right) & = & 
\KM \cdot \lbrace (v_L R_{fi}^{L} + v_T R_{fi}^{T} + v_{TT}R_{fi}^{TT} + 
                    v_{LT} R_{fi}^{LT}) + \nonumber \\
 & & \qquad \qquad + h (v_{T'}R_{fi}^{T'} + v_{LT'} R_{fi}^{LT'}) \rbrace
\eea
with
\begin{equation}
K_{Mott} = \frac{M_p m_\pi p_p^{cm}}{8\pi^3 W} \sigma_{Mott}\,\mbox{.}
\end{equation}
$W$  is the invariant mass of the recoiling hadronic system,
$p_p^{cm}$  the proton momentum in the center-of-momentum frame, 
and $M_p$  and $m_\pi$  are the proton and pion rest mass, respectively.
The electron kinematics enters into the factors $v_M$
($M=L,T,TT,LT,T',LT'$):
\bea \label{eq:v}
v_L    & = & \left( \frac{Q^2}{\vec q\,^2} \right)^2 \cdot 
             \left( \frac{W}{M_p} \right)^2                     \nonumber \\
v_T    & = & \frac{1}{2}\left( \frac{Q^2}{\vec q\,^2} \right) +
             \tan^2\frac{\vartheta_e}{2}                      \nonumber \\
v_{TT} & = & - \frac{1}{2}\left( \frac{Q^2}{\vec q\,^2} \right) \nonumber \\
v_{LT} & = & - \frac{1}{\sqrt{2}} \left( \frac{Q^2}{\vec q\,^2} \right)
             \sqrt{\left( \frac{Q^2}{\vec q\,^2} \right)+
                   \tan^2{\frac{\vartheta_e}{2}}} \cdot 
                   \frac{W}{M_p}                                \nonumber \\
v_{T'} & = & \sqrt{\left( \frac{Q^2}{\vec q\,^2} \right)+
                   \tan^2{\frac{\vartheta_e}{2}}} \tan{\frac{\vartheta_e}{2}} 
                                                            \nonumber \\
v_{LT'}& = & - \frac{1}{\sqrt{2}} \left( \frac{Q^2}{\vec q\,^2} \right)
             \tan{\frac{\vartheta_e}{2}} \cdot \frac{W}{M_p} 
\eea
In the above equations, $\vartheta_e$  is the electron scattering angle,
$\vec q\,^2$  the square of the three-momentum transfer,
$Q^2 = 4EE'\sin^2(\vartheta_e/2)$  is the negative squared four-momentum 
transfer,
and $h$  is the longitudinal polarization of the electron beam.
The structure of the %$p$--$\pi^0$  
hadronic system is contained in the six structure
functions $R_{fi}^M$, which
implicitly contain the proton polarization.
The dependence on
proton polarization can be made explicit, leading to a total of 18 structure 
functions \cite{RD,DT} in the cross section: 
\bea \label{eq:x-sec18}
\left( \frac{d\sigma}{dE'\Omega_e\Omega_p^{cm}} \right) & = & 
      K_{Mott} \cdot \Big\{ v_L(\RL+\Pi_n\RLn) + v_T(\RT+\Pi_n\RTn) + \nonumber 
\\ 
 & &      \quad   v_{LT}\left[ (\RLT+\Pi_n\RLTn) \cos\Phi +
                               (\Pi_ll\RLTl+\Pi_t\RLTt) \sin\Phi \right] +
\nonumber\\
 & &      \quad   v_{TT}\left[ (\RTT+\Pi_n\RTTn) \cos{2\Phi} +
                               (\Pi_l\RTTl+\Pi_t\RTTt) \sin{2\Phi} \right] +
\nonumber\\
 & &      \quad   h \cdot \Big\{
                  v_{LT'}\left[ (\RLTp+\Pi_n\RLTpn) \sin\Phi +
                               (\Pi_l\RLTpl+\Pi_t\RLTpt) \cos\Phi \right] +
\nonumber\\
 & &      \qquad  v_{T'} \left[ \Pi_l\RTTpl + \Pi_t\RTTpt \right]  \Big\}
           \Big\}   
\eea
$\Pi_{n,l,t} = \pm 1$ are the
projections of the proton spin in its rest frame onto the axes $n$, $l$, $t$
depicted in Fig.\ref{fig:kin}.
The longitudinal unit vector, $\hat l$,  is in the direction of the 
proton momentum in the center-of-momentum frame, 
$\hat n = \hat q \times \hat l / \sin{\tcm}$  
points normal to the reaction plane 
and $\hat t = \hat n \times \hat l$  
is perpendicular to the proton momentum in
the reaction plane. 
The connection between the R structure functions and the W structure functions 
as defined by Raskin and Donnelly \cite{RD} is given in the appendix.

From the cross section of Eq.(\ref{eq:x-sec18}) one gets for 
the recoil proton polarization components
\bea
\label{eq:pltn_cm}
\sigma_0 P_l &=& \KM \left\{ 
   v_{LT}\RLTl \sin\Phi + v_{TT}\RTTl \sin2\Phi + 
   h \left[ v_{LT'}\RLTpl \cos\Phi + v_{T'}\RTTpl \right] \right\}  
   \nonumber \\
\sigma_0 P_t &=& \KM \left\{ 
   v_{LT}\RLTt \sin\Phi + v_{TT}\RTTt \sin2\Phi + 
   h \left[ v_{LT'}\RLTpt \cos\Phi + v_{T'}\RTTpt \right] \right\}  \\
\sigma_0 P_n &=& \KM \left\{
   v_L\RLn + v_T\RTn + v_{LT}\RLTn \cos\Phi + v_{TT}\RTTn \cos2\Phi +
   h v_{LT'}\RLTpn \sin\Phi   \right\} ,
   \nonumber
\eea
where $\sigma_0$  represents the proton polarization independent part
of the cross section.
The recoil proton polarization can be 
split into the electron polarization dependent part (transferred polarization),
$P_{\{n,l,t\}}'$,
which is proportional to $h$,
and an electron polarization independent induced polarization.
From the above equations  
the transferred polarization components are given by:
\bea
\label{eq:Pt_simple}
\sigma_0 P_t' &=& h \cdot \KM \cdot 
                  \left[ v_{T'} \RTTpt + v_{LT'} \RLTpt\cos{\Phi} \right]
\nonumber \\
\sigma_0 P_l' &=& h \cdot \KM \cdot 
                  \left[ v_{T'} \RTTpl + v_{LT'} \RLTpl\cos{\Phi} \right] \\
\sigma_0 P_n' &=& h \cdot \KM \cdot 
                  v_{LT'} \RLTpn\sin{\Phi} \nonumber 
\eea
There are two terms contributing to the polarization 
component $P_t'$.
The first one is independent of $\Phi$ and points
always into $t$  direction of the reaction plane 
reference frame, which rotates with the out-of-plane angle $\Phi$
(see Fig.\ref{fig:kin}). 
Viewed from the electron scattering plane, the polarization related to
this term points into opposite directions left ($\Phi = 0$) 
and right ($\Phi = \pi$) of $\vec q$ 
and therefore vanishes in the case of parallel kinematics
$\tcm = 0$.
Correspondingly, \RTTpt \, carries an implicit 
$\sin\tcm$-dependence.
The second term depends on $\cos\Phi$, like the projection 
of a polarization which is fixed in the electron scattering plane onto the 
rotating $\{n, l, t \}$ frame.
This part {\em does not} vanish in parallel kinematics.

Similarly the other components of Eq.(\ref{eq:Pt_simple}) also contain 
projections of a fixed polarization  in the electron scattering plane.
The natural choice for a polarization, $P$, that is fixed in the 
electron scattering plane is the $\{x,y,z\}$  frame of Fig.\ref{fig:kin},
which is related to the $\{n,l,t\}$ system by a simple rotation:
\bea
\label{eq:transformation}
P_x &=& \Pl \sin\tcm \cos\Phi + \Pt \cos\tcm \cos\Phi - \Pn \sin\Phi  
\nonumber \\
P_y &=& \Pl \sin\tcm \sin\Phi + \Pt \cos\tcm \sin\Phi + \Pn \cos\Phi  \\
P_z &=& \Pl \cos\tcm - \Pt \sin\tcm
\nonumber 
\eea
In the case of parallel kinematics this transformation remains still defined.
The angle $\Phi$  then plays the role of the orientation of the transverse
polarization, $P_t$,  relative to the electron scattering plane.

The proton polarization components can be expressed
by the multipole decomposition of the structure functions according to
\cite{RD}.
Restricting the expansion in the usual way to s and p waves and
retaining only terms with the dominant $\mep$  amplitude,
one gets for the case of strictly parallel kinematics
with $\tcm = 0$:
\bea
\sigma_0 P_x &=& h \cdot \KLTp \cdot \sqrt{2} \cdot \Re \left\{
                 \snp\mep + \sem\mep + 4 \sep\mep  \right\}  
                 \label{eq:px_simple} \\
\sigma_0 P_y &=& - \KLT \cdot \sqrt{2} \cdot \Im \left\{
                 \snp\mep + \sem\mep + 4 \sep\mep  \right\}  
                 \label{eq:py_simple} \\
\sigma_0 P_z &=& h \cdot \KTp \cdot \left[ |\mep|^2 +
                 \Re \left\{ 6\eep\mep^*+2\mep(\enp^*-\mem^*) \right\} \right]
                 \label{eq:pz_simple} 
\eea  
with
\begin{equation}
K_M   = K_{Mott} \cdot v_{M} \cdot \frac{4 \pi W^2}{\alpha m_\pi M_p^2}
~;~~~ M = LT', LT, T' \mbox{.} 
\end{equation}

The two in-plane components, $P_x$  and $P_z$, are proportional to
the electron helicity, $h$, and vanish with unpolarized electron beam.
Contrary, the component normal to the electron scattering plane, $P_y$,  
is independent of $h$  and thus shows up
already with {\em unpolarized} beam.

$P_x$  carries in parallel kinematics a high sensitivity to the
small longitudinal quadrupole amplitude, $S_{1+}$, due to the interference with
the large $\mep$  amplitude.
%The real part of Eq.(\ref{eq:px_simple}) 
It is, however, not solely determined by resonant amplitudes, 
but receives both resonant and non-resonant contributions.
The induced polarization, $P_y$  (Eq.\ref{eq:py_simple}), 
measures the imaginary part of the same combinations of interference terms
of which $P_x$ (Eq.\ref{eq:px_simple}) determines the real part.
This offers the possibility to disentangle resonant and non-resonant
pieces, which will later be discussed in more detail.
$P_z$  is dominated by $|\mep|^2$. 
Therefore, the ratio of the two in-plane polarization
components, $P_x / P_z$,  is directly related to $S_{1+} / \mep$. 
%These two polarization components
$P_x$ and $P_z$ 
are simultaneously accessible behind a
(spin precessing) magnetic system like a proton spectrometer.
In the ratio $P_x/P_z$  the absolute values of both the electron
beam polarization and the analyzing power of the proton polarimeter cancel out,
which otherwise represent major sources of systematic uncertainties.

With real detectors the polarization components are averaged over
finite acceptances around parallel kinematics.
This will be discussed in the next section along with the influence of the 
non-leading terms in the s and p wave approximation.

\subsection{Polarization observables in the laboratory frame}
\label{sec:angle}

The considerations of the preceeding section illustrate the sensitivity 
of the recoil proton polarization
to the $S_{1+}$ quadru\-pole amplitude for parallel kinematics.
A real experiment will cover a finite solid angle around the
strictly parallel case. 
Therefore, in this section the azimuthal averaging of the polarization 
components $P_{x,y,z}^{lab}$  is considered.
For this discussion the polarization (Eq.(\ref{eq:pltn_cm}))
is projected from the center-of-momentum into the laboratory frame
\cite{Gross89}.
The corresponding transformation is given by the so-called
Wigner-rotation \cite{Wigner-rotation}:
\bea
\label{eq:wigner_rot}
P_t^{lab} &=& P_t \cos{\vartheta_W} + P_l \sin{\vartheta_W}  \nonumber \\
P_l^{lab} &=& - P_t \sin{\vartheta_W} + P_l \cos{\vartheta_W}          \\
P_n^{lab} &=& P_n                                            \nonumber
\eea
The Wigner angle, $\vartheta_W$, is given by
\begin{equation}
\sin{\vartheta_W} = \frac{1+\gamma}{\gamma^{cm}+\gamma^{lab}} \cdot
                    \sin(\tcm-\theta_p^{lab}),
\end{equation}
where the Lorentz factors $\gamma$, $\gamma^{cm}$  and $\gamma^{lab}$
are related to the velocities of the center-of-momentum frame against the
laboratory frame, and of the proton in the cm and lab frames, respectively.
The transformation
\bea
\label{eq:xyz_projection}
P_x^{lab} &=& \Pl^{lab} \sin\theta_p^{lab} \cos\Phi + 
              \Pt^{lab} \cos\theta_p^{lab} \cos\Phi - 
              \Pn^{lab} \sin\Phi                       \nonumber \\
P_y^{lab} &=& \Pl^{lab} \sin\theta_p^{lab} \sin\Phi + 
              \Pt^{lab} \cos\theta_p^{lab} \sin\Phi + 
              \Pn^{lab} \cos\Phi                                 \\
P_z^{lab} &=& \Pl^{lab} \cos\theta_p^{lab} -  
              \Pt^{lab} \sin\theta_p^{lab}             \nonumber 
\eea
projects the polarization as seen in the laboratory reaction plane 
(Eq.(\ref{eq:wigner_rot})) onto the \{x,y,z\}-frame related to the 
electron scattering plane.
The \{x,y,z\}-components of Eq.(\ref{eq:xyz_projection}) are 
azimuthally averaged around the direction of the
momentum transfer, $\vec q$,
which is indicated by the bar in the following equations.
Only those terms with even powers of $\sin\Phi$ and $\cos\Phi$ 
survive the integration over $\Phi$. 
Keeping for the sake of clarity only terms containing the dominant $\mep$
amplitude, the result is:
\bea
\label{eq:Px_lab_M1+}
(\overline{\sigma_0 P_x})^{lab} &=& h\cdot\KLTp\cdot\sqrt{2}\cdot\Re \big\{
         \snp\mep\cdot\frac{1}{2}\big[
         -4\sin\tcm(\sin\tlab\cos\tw + \cos\tlab\sin\tw)       \nonumber \\
    & &  \qquad 
         +\cos\tcm(\cos\tlab\cos\tw - \sin\tlab\sin\tw + 1) 
                                        \big] \quad +          \nonumber \\
    & &  
         \sem\mep\cdot\frac{1}{2}\big[
         1+(2-\cos^2\tcm)(\cos\tlab\cos\tw - \sin\tlab\sin\tw)
                                                               \nonumber \\
    & &  \qquad
         -\sin\tcm\cos\tcm(\cos\tlab\sin\tw + \sin\tlab\cos\tw)
                                        \big] \quad +          \nonumber \\
    & &  
         \sep\mep\cdot\frac{1}{2}\big[
         4(2\cos^2\tcm-1)(\cos\tlab\cos\tw-\sin\tlab\sin\tw)   
                                                               \nonumber \\
    & &  \qquad
         -10\cos\tcm\sin\tcm(\cos\tlab\sin\tw+\sin\tlab\cos\tw) 
                                                               \nonumber \\
    & &  \qquad
         + 2(3\cos^2\tcm-1)
                                        \big]
                                                                     \big\} \\
\label{eq:Py_lab_M1+}
(\overline{\sigma_0 P_y})^{lab} &=& - \KLT\cdot\sqrt{2}\cdot\Im \big\{
         \snp\mep\cdot\frac{1}{2}\big[
         -4\sin\tcm(\sin\tlab\cos\tw + \cos\tlab\sin\tw)       \nonumber \\
    & &  \qquad 
         +\cos\tcm(\cos\tlab\cos\tw - \sin\tlab\sin\tw + 1) 
                                        \big] \quad +          \nonumber \\
    & &  
         \sem\mep\cdot\frac{1}{2}\big[
         1+(2-\cos^2\tcm)(\cos\tlab\cos\tw - \sin\tlab\sin\tw)
                                                               \nonumber \\
    & &  \qquad
         -\sin\tcm\cos\tcm(\cos\tlab\sin\tw + \sin\tlab\cos\tw)
                                        \big] \quad +          \nonumber \\
    & &  
         \sep\mep\cdot\frac{1}{2}\big[
         4(2\cos^2\tcm-1)(\cos\tlab\cos\tw-\sin\tlab\sin\tw)   
                                                               \nonumber \\
    & &  \qquad
         -10\cos\tcm\sin\tcm(\cos\tlab\sin\tw+\sin\tlab\cos\tw) 
                                                               \nonumber \\
    & &  \qquad
         + 2(3\cos^2\tcm-1)
                                        \big]
                                                                     \big\} \\
\label{eq:Pz_lab_M1+}
(\overline{\sigma_0 P_z})^{lab} &=& h\cdot\KTp\cdot \big\{
         |\mep|^2 \big[ (2-\cos^2\tcm)(\cos\tlab\cos\tw-\sin\tlab\sin\tw)
                                                               \nonumber \\
    & &  \qquad         - \sin\tcm\cos\tcm(\cos\tlab\sin\tw+\sin\tlab\cos\tw)
                  \big]  \quad +
                                                               \nonumber \\
    & &  \Re\{6\mep^*\eep\} 
         \big[ (2\cos^2\tcm-1)(\cos\tlab\cos\tw-\sin\tlab\sin\tw)
                                                               \nonumber \\
    & &  \qquad - \sin\tcm\cos\tcm(\cos\tlab\sin\tw+\sin\tlab\cos\tw) 
         \big] \quad +
                                                               \nonumber \\
    & &  \Re\{\mem^*\mep\}
         \big[ (-1-\cos^2\tcm)(\cos\tlab\cos\tw-\sin\tlab\sin\tw)
                                                               \nonumber \\
    & &  \qquad - \sin\tcm\cos\tcm(\cos\tlab\sin\tw+\sin\tlab\cos\tw)
         \big] \quad +
                                                               \nonumber \\
    & &  \Re\{\enp^*\mep\}
         \big[ 2\cos\tcm(cos\tlab\cos\tw - \sin\tlab\sin\tw)  \nonumber \\
    & &  \qquad - \sin\tcm(\cos\tlab\sin\tw + \sin\tlab\cos\tw)
         \big]
                                                        \big\}
\eea

\begin{figure}
\begin{center}
\epsfig{file=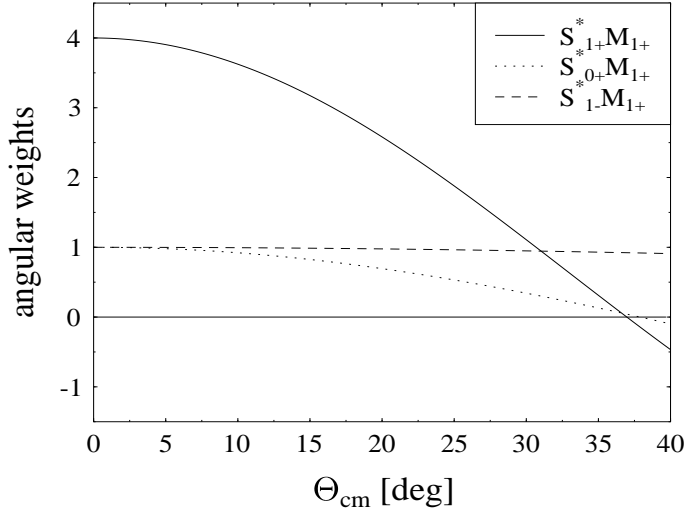,width=11cm,height=8cm}
\caption{ Relative angular weights of the leading 
         multipole terms of $P_x$ and $P_y$,
         which are the same for both components (see text),
         for the kinematics described in section \ref{sec:rates}.}
\label{fig:weights_x_lab}
\end{center}
\end{figure}
The angular coefficients of the interference terms are the same in
Eq.(\ref{eq:Px_lab_M1+}) and (\ref{eq:Py_lab_M1+}).
They are plotted in Fig.\ref{fig:weights_x_lab}. 
% The coefficients are identical for P_x^{lab}$ and $P_y^{lab}$.
The sensitivity to the $\sep\mep$ interference term decreases with increasing
$\tcm$.
$\snp\mep$ shows practically the same behaviour, but reduced by a factor 4,
while the weight of $\sem\mep$ is almost constant.

In the limit of parallel kinematics, 
Eqs.(\ref{eq:Px_lab_M1+}-\ref{eq:Pz_lab_M1+}) reduce to
Eqs.(\ref{eq:px_simple}-\ref{eq:pz_simple}).
Keeping also the non-leading terms in the s and p wave approximation, 
one arrives at: 
\bea
\label{eq:px_par_complete}
(\overline{\sigma_0 P_x})^{lab}_{\theta \rightarrow 0}   &=& 
                \KLTp\cdot h\cdot\sqrt{2} \cdot \Re \big\{
                (\snp+\sem+4\sep)(\enp+3\eep+\mep-\mem)     \big\}~~~\\
\label{eq:py_par_complete}
(\overline{\sigma_0 P_y})^{lab}_{\theta \rightarrow 0}   &=& 
                - \KLT \cdot \sqrt{2} \cdot \Im \big\{
                (\snp+\sem+4\sep)(\enp+3\eep+\mep-\mem)     \big\}~~~\\
\label{eq:pz_par_complete}
(\overline{\sigma_0 P_z})^{lab}_{\theta \rightarrow 0}   &=& 
                \KTp\cdot h\cdot \Big[
                |\enp|^2+9|\eep|^2+|\mem|^2+|\mep|^2 \nonumber \\
 & & \quad +\Re \left\{ 2\enp^*(6\eep+\mep-\mem)+6\mep^*\eep 
                         - 2 \mem^*\mep                       \right\}\Big].
\eea
Thus, in parallel kinematics, $P_y$ contains the imaginary part of the same 
interference terms as the real part in $P_x$. 
This fact can be exploited for a separation of contributions due to the
Delta-resonance from other contributions, which are caused either by 
non-resonant $\pi^0$-production or by higher nucleon resonances. 

\subsection{Separation of resonant and non-resonant pieces}
\label{sec:separation}

The multipole amplitudes of 
Eqs.(\ref{eq:px_par_complete}-\ref{eq:pz_par_complete})
are not solely determined by the $\Delta$-resonance, 
but contain both resonant and non-resonant pieces.
Therefore, in the following, the multipole combinations of $P_x$\,and $P_y$
(Eqs.\ref{eq:px_par_complete} and \ref{eq:py_par_complete})
are split into their resonant and non-resonant parts.
This is closely related to the decomposition of the physical 
$\pi^0$-electroproduction amplitudes, $A_i^{\pi^0}$, into isospin
$\frac{1}{2}$  and $\frac{3}{2}$  channels \cite{Komar89}.
\begin{equation}
A_i^{\pi^0} = A_i^{1/2} + \frac{2}{3} A_i^{3/2}
\hspace{0.3cm} ; \hspace{1.0cm} A = M,\,E,\,S
\end{equation}
As stated by the Watson Final State Theorem \cite{Watson},
all $A_{1+}^{3/2}$  amplitudes show the almost purely resonant behaviour
of $\mep^{3/2}$.
All other multipoles are considered as non-resonant.
\bea
\snp+\sem+4\sep &=& [\snpx^{1/2}+\semx^{1/2}+4\sepx^{1/2}+
                     \frac{2}{3}(\snpx^{3/2}+\semx^{3/2})]^* 
                   + \frac{8}{3} (\sepx^{3/2})^*   \nonumber    \\
                &=& [S_{non}^*] + S_{res}^*           \\
\enp+3\eep+\mep-\mem &=& [\enp^{1/2}+3\eep^{1/2}+\mep^{1/2}-\mem^{1/2}+
                          \frac{2}{3}(\enp^{3/2}-\mem^{3/2})] 
                          \nonumber \\
   & &    \quad          + \frac{2}{3}(3\eep^{3/2}+\mep^{3/2}) \nonumber \\
                     &=& [E,M_{non}] + E,M_{res}\mbox{.}
\eea
If, at the position of the $\Delta$ resonance, 
all terms without the by far dominating $\Im\{\mep^{3/2}\}$ 
are neglected,
Eqs.(\ref{eq:px_par_complete}) and (\ref{eq:py_par_complete}) 
can be written as
\bea
\label{eq:px_M1+_3/2}
(\overline{\sigma_0 P_x})^{lab}_{\theta \rightarrow 0}   &=&
       \KLTp\cdot h\cdot\sqrt{2} \cdot 
       \Im\{S_{non}^*+S_{res}^*\} \cdot \frac{2}{3} \Im\{\mep^{3/2}\}      \\
\label{eq:py_M1+_3/2}
(\overline{\sigma_0 P_y})^{lab}_{\theta \rightarrow 0}   &=&
       - \KLT \cdot \sqrt{2} \cdot
       \Re\{S_{non}^*\} \cdot \frac{2}{3} \Im\{\mep^{3/2}\}
\eea

The real parts of resonant amplitudes vanish directly on top of the
resonance and therefore the corresponding terms do not occur in 
the above equations.
According to Eqs.(\ref{eq:px_M1+_3/2}) and (\ref{eq:py_M1+_3/2})
$P_x$  measures the sum of the resonant longitudinal 
quadrupole component, 
$S_{res}^* = \frac{8}{3} (S_{1+}^{3/2})^*$,  
and nonresonant contributions, $S_{non}^*$,
whereas $P_y$  is solely sensitive to $S_{non}^*$.
In the (hypothetical) case of a single, pure resonance
where all real parts vanish on top of the resonance, $P_y$  would thus
be identical zero.

However, purely real Born terms 
$\snpx^{1/2,3/2}$, $\semx^{1/2,3/2}$, and $\sepx^{1/2}$  
result already in a nonvanishing $P_y$.
On the other hand, for real Born terms $\Im\{S_{non}\}$ vanishes, i.e.
Eq.(\ref{eq:px_M1+_3/2}) yields:
\begin{equation}
\label{eq:px_isodecomp}
(\overline{\sigma_0 P_x})^{lab}_{\theta \rightarrow 0}   =
       \KLTp\cdot h\cdot\sqrt{2} \cdot \frac{16}{9} \cdot 
       \Im\{(\sepx^{3/2})^*\} \cdot \Im\{\mep^{3/2}\}
\end{equation}
This means that, within the approximations discussed,
$P_x$  contains directly the wanted isospin 3/2 part of the $\sepx$  amplitude.

Non-Born contributions might occur due to 
either rescattering processes or
higher resonances,
like $\semx^{1/2}$  from the Roper N(1440).
If there were non-Born imaginary parts contributing, 
Eq.(\ref{eq:px_isodecomp}) would be more complicated.
Such terms are in principle detectable through $P_y$, 
because real and imaginary parts of the amplitudes are related by 
fixed phases as requested by Watson's Final State Theorem.
Therefore imaginary parts in $S_{non}$ %of Eq.(\ref{eq:px_M1+_3/2})
go along with an altered 
real part $\Re\{S_{non}\}$ %in Eq.(\ref{eq:py_M1+_3/2})
as compared to purely real non-resonant amplitudes.
\section{Experimental aspects}
\label{sec:experimentals}

The polarization of recoil protons can be measured in a focal plane polarimeter
behind a magnetic spectrometer, like the proton polarimeter \cite{Pospisch}
of the A1 collaboration \cite{A1} at MAMI. 
Such a device measures the azimuthal asymmetry of protons which were
inclusively scattered in a carbon secondary scatterer.
With this process, only the two polarization components 
perpendicular to the proton momentum are accessible.
Due to the spin precession in the spectrometer magnetic system, 
these two polarization components measured in the focal plane
are linear combinations of all three components 
at the target, $P_x, P_y, P_z$.
Provided a complete understanding of the spin precession,
the measurement of only two focal plane polarization components is
nevertheless sufficient to determine all three target components,
because there is additional information from flipping the
electron beam helicity:
$P_x$  and $P_z$  are odd under helicity reversal, while $P_y$  is even
(cf. Eqs.(\ref{eq:px_simple}-\ref{eq:pz_simple})).

The averaging over the azimuthal angle, $\Phi$,
which leads to the expressions discussed in section \ref{sec:angle},
can be easily accomplished in the case of parallel kinematics where 
the spectrometer sits in the momentum transfer direction. 
Here the sensitivity to the longitudinal quadrupole amplitude, $S_{1+}$,  
is maximum. 
It is higher than in previously proposed experiments with distinct
measurements left and right of the momentum transfer direction
\cite{Lourie90, Bates, MAMI-prop}.
The comparatively high degree of proton polarization in those experiments 
is only due to the mixing of the large $P_z$  component,
which according to Eq.(\ref{eq:Pz_lab_M1+}) contains a $|\mep|^2$  term, 
into the considered $P_t$  polarization components at finite angles \tcm.

In contrast to a non-magnetic polarimeter, where the longitudinal 
proton polarization component is inaccessible,  
$P_x$  and $P_z$  can be measured simultaneously behind the spectrometer.
This allows the mesaurement of the ratio $P_x/P_z$
with obvious advantages:
\begin{enumerate}
\item
The leading term of this ratio is directly 
$\Re\{\sep\mep\}/|\mep|^2$.
\item
In the polarization ratio the absolute value of the electron beam 
polarization cancels out.
\item
The recoil polarizations are determined by polarimeter 
asymmetries with a common effective analyzing power.
The polarization ratio is thus also independent of
the absolute value of the polarimeter's analyzing power.  
\end{enumerate}
Therefore such a measurement can be performed without monitoring the 
electron beam polarization.
The beam polarization need even not be constant over time, because both
recoil polarization components are measured truely simultaneously.
The absolute calibration of the effective polarimeter analyzing 
power is neither required, since in the ratio it cancels out, too.
A similar polarization-ratio method was successfully employed in a recent 
measurement of the neutron electric formfactor \cite{FKL96,MO97}.

The influence of possible non-Born contributions to the 
measured ratio can be studied via the induced polarization, $P_y$. 
This component is independent of the electron beam polarization and thus
more sensitive to false systematic asymmetries.
For the analysis of $P_y$  the absolute calibration 
of the proton polarimeter is therefore desirable,
although a ratio measurement $P_y/P_z$  could also be imagined.
In any case, the beam polarization must be known,
since in the $P_y/P_z$-ratio the polarimeter analyzing power cancels, 
but the beam polarization does not.

\subsection{Expected proton polarizations in a realistic experiment}
\label{sec:rates}

Accounting only for the leading terms in the above expressions
(cf. Eqs.(\ref{eq:px_simple},\ref{eq:pz_simple}))
and neglecting a possible offset due to imaginary parts of the 
non-resonant amplitudes,
the recoil proton polarization in parallel kinematics can be estimated by
\bea
P_x &=& \frac{1}{\sigma_0} h \KM v_{LT'} N^2 \sqrt{2} \Re\{4\sep\mep\} \\
P_z &=& \frac{1}{\sigma_0} h \KM v_{T'} N^2 |\mep|^2 \mbox{.}
\eea
With the proton polarization independent cross section approximated through
\begin{equation}
\sigma_0 = \KM v_T N^2 |\mep|^2 \mbox{,}
\end{equation}
one receives
\bea
P_x &=& 4 \sqrt{2} h \frac{v_{LT'}}{v_{T}} \frac{\Re\{\sep\mep\}}{|\mep|^2}
\nonumber \\
\label{eq:Px_approx}
    &=& -8 h \frac{\tan (\vartheta_e/2)}{1+\frac{2\vec q\,^2}{Q^2}
            \tan^2(\vartheta_e/2)} \cdot \frac{W}{M_p} \cdot 
            \frac{\Re\{\sep\mep\}}{|\mep|^2} 
\nonumber \\
    &=& -8\cdot h \cdot \epsilon \cdot \tan(\vartheta_e/2) \cdot \frac{W}{M_p} 
          \cdot \frac{\Re\{\sep\mep\}}{|\mep|^2}\,,\\
P_z &=& h \frac{v_{T'}}{v_{T}}\frac{|\mep|^2}{|\mep|^2} 
\nonumber \\
\label{eq:Pz_approx}
    &=& h \sqrt{1 - \epsilon^2} \,. 
\eea
$\epsilon = [ 1 + (2 \vec q\,^2/Q^2)\tan^2\frac{\vartheta_e}{2} ]^{-1}$
is the virtual photon's degree of transverse polarization.
Making use of the relations of appendix D of \cite{DT}
between CGLN amplitudes \cite{CGLN} and structure functions,
the above relation for
$P_z$  can be shown to hold generally in parallel kinematics, 
i.e. independently of the approximations discussed.
Fixed by kinematical variables only, $P_z$
might thus be used for calibration checks.
Applying the electron kinematics of the MAMI $N$-$\Delta$  proposal
\cite{MAMI-prop},
$ E            = 0.855\,\mbox{GeV}, \,
  W            = 1.232\,\mbox{GeV}, \,
  Q^2          = 0.12 \,\mbox{(GeV/c)}^2, \,
  \vartheta_e  = 32^o , \,
  |\vec q|     = 0.53 \,\mbox{GeV} \, 
$,
Eqs.(\ref{eq:Px_approx}) and (\ref{eq:Pz_approx}) yield
\bea
\label{eq:px_estimate}
P_x &=& - 2.2 \cdot h \cdot 
               \frac{\Re\{\sep\mep\}}{|\mep|^2}  \\
P_z &=&   0.7 \cdot h \,.
\eea
Thus, a quadrupole contribution of the order of $-5$\,\%
causes a transverse proton polarization of $P_x \simeq 7.6\,\%$ with
an electron  beam polarization of 70\,\%,
which now routinely is achieved \cite{Aule97}.
The longitudinal proton polarization then is $P_z = 49\,\%$.

\section{Summary and conclusion}
\label{sec:conclusion}

The $p (\vec e, e' \vec p)\pi^0$  reaction with measurement of the
recoil proton polarization has a large potential towards the precise 
determination of the longitudinal quadrupole component, $\sepx$, 
in the N to $\Delta$  transition.
In particular in parallel kinematics,
it offers on top of the $\Delta$  resonance a high sensitivity to the 
$\sep\mep$  interference term. 
This is clearly revealed when the process is discussed
in the appropriate $\{x,y,z\}$  coordinate frame,
which is fixed to the electron scattering plane (see Fig.\ref{fig:kin}).
Here the polarization transfer from the electron takes a simple
form and is not obscured by projections onto rotating reference frames.

The ratio of the recoil proton polarization components, $P_x/P_z$,  
is directly related to $\Re\{\sep\mep\}/|\mep|^2$.
If both components are measured simultaneously after the deflection
in a magnetic spectrometer, 
the absolute values of both electron beam polarization and polarimeter
analyzing power cancel out. 
Therefore small experimental uncertainties can be achieved.
The electron beam helicity independent polarization component,
$P_y$,  offers the opportunity to determine possible non-Born contributions.

\section{Acknowledgements}

I thank R. Beck, J. Friedrich, F. Klein and L. Tiator for many 
fruitful discussions.
This work was supported by the Deutsche Forschungsgemeinschaft (SFB201).

\newpage
{\Large {\bf Appendix}} \\[.3cm]

The relation between the R and W structure functions of
\cite{RD} are explicitly:
\begin{alignat}{4}
\hat{R}_L & = W^L(0)       & \qquad \hat{R}_L^n    & = W^L(n)   
                           & \qquad \hat{R}_T      & = W^T(0)   
                           & \qquad \hat{R}_T^n    & = W^T(n)   
\nonumber \\
\hat{R}_{TT} & = W^{TT}(0) & \qquad \hat{R}_{TT}^n & = W^{TT}(n)
                           & \qquad \hat{R}_{TT}^l & = \tilde{W}^{TT}(l) 
                           & \qquad \hat{R}_{TT}^t & = \tilde{W}^{TT}(s)
\nonumber \\
\hat{R}_{LT} & = W^{TL}(0) & \qquad \hat{R}_{LT}^n & = W^{TL}(n)
                           & \qquad \hat{R}_{LT}^l & = \tilde{W}^{TL}(l) 
                           & \qquad \hat{R}_{LT}^t & = \tilde{W}^{TL}(s)      
\nonumber \\
\hat{R}_{LT'}& =\tilde{W}^{TL'}(0) 
                           & \qquad \hat{R}_{LT'}^n& = \tilde{W}^{TL'}(n)
                           & \qquad \hat{R}_{LT'}^l& = W^{TL'}(l) 
                           & \qquad \hat{R}_{LT'}^t& = W^{TL'}(s)  
\nonumber \\
& & &                      & \qquad \hat{R}_{TT'}^l& = W^{T'}(l) 
                           & \qquad \hat{R}_{TT'}^t& = W^{T'}(s)
\nonumber 
\end{alignat} 
with
\[ \hat R_M = \frac{R_M}{N^2}\qquad \mbox{and}\qquad 
   N^2 = \frac{4 \pi W^2}{\alpha m_{\pi} M_N^2}. \]

\end{document}